\documentclass[sn-nature]{sn-jnl}

\usepackage{graphicx}%
\usepackage{multirow}%
\usepackage{amsmath,amssymb,amsfonts}%
\usepackage{amsthm}%
\usepackage{mathrsfs}%
\usepackage[title]{appendix}%
\usepackage{xcolor}%
\usepackage{textcomp}%
\usepackage{manyfoot}%
\usepackage{booktabs}%
\usepackage{algorithm}%
\usepackage{algorithmicx}%
\usepackage{algpseudocode}%
\usepackage{listings}%

\usepackage[normalem]{ulem}
\usepackage{xcolor}
\usepackage{comment}
\newcounter{comments}
\definecolor{Red}{rgb}{0.8,0,0}
\definecolor{Green}{rgb}{0.2,0.6,0.2}

\newcommand{\Ff}{{\mathbb F}}

\begin{document}

\title[]{Canalization as a stabilizing principle of gene regulatory networks: a discrete dynamical systems perspective}


\author*[1]{\fnm{Claus} \sur{Kadelka}}\email{ckadelka@iastate.edu}

\affil*[1]{\orgdiv{Department of Mathematics}, \orgname{Iowa State University}, \orgaddress{\street{411 Morrill Rd}, \city{Ames}, \postcode{50011}, \state{IA}, \country{United States}}}


\abstract{
Gene regulatory networks exhibit remarkable stability, maintaining functional phenotypes despite genetic and environmental perturbations. Discrete dynamical models, such as Boolean networks, provide systems biologists with a tractable framework to explore the mathematical underpinnings of this robustness. A key mechanism conferring stability is canalization. This perspective synthesizes historical insights, formal definitions of canalization in discrete dynamical models, quantitative measures of stability, illustrative applications, and emerging challenges at the interface of theory and experiment.
} 




\maketitle

\section{Introduction}\label{sec:intro}
Gene regulatory networks (GRNs) orchestrate cellular behavior by determining 
which genes are expressed, when, and to what extent. Through cascades of 
regulatory interaction -- where transcription factors bind promoters, miRNAs 
silence transcripts, and proteins modulate each other's activity -- GRNs 
translate genomic information into functional phenotypes. The key step of gene regulation is an inherently stochastic process~\cite{Elowitz02}, and the numbers of intra- and extra-cellular regulatory signals fluctuate widely~\cite{kaern2005stochasticity}.
Thus, understanding how GRNs perform particular functions and do so consistently in the face of ubiquitous variability constitutes a fundamental biological question.
This robustness is puzzling: if gene expression is stochastic and cellular 
signals fluctuate, how do GRNs reliably produce consistent phenotypes? 
The answer lies in their architecture. Numerous putative ``design principles" have been proposed, most of which describe how certain aspects of GRNs differ from random networks~\cite{shen2002network,macneil2011gene,deritei2016principles,gorochowski2018organization,daniels2018criticality, kadelka2024meta, kadelka2024canalization}. However, a comprehensive understanding of the mechanistic principles underlying the structure of GRNs and how they collectively contribute to robust phenotypes is still lacking. Among these principles, canalization stands out as both historically foundational 
and mathematically tractable, making it an ideal lens through which to understand 
GRN robustness. In this perspective, I synthesize recent mathematical advances that quantify canalization and highlight open challenges in connecting these measures to biological function.

The concept of canalization in gene regulation dates back to the work of geneticist Conrad Waddington in the 1940s~\cite{Wad}, who proposed it as a potential explanation for how embryonic development can reliably produce predictable phenotypes despite substantial environmental variation and frequent genetic mutations. 
More broadly, canalization describes the capacity of a gene regulatory program to maintain stability in the face of diverse perturbations. By buffering against the deleterious effects of mutations, canalization permits the accumulation of genotypic variation without corresponding phenotypic change~\cite{gibson2000canalization}. 
When extreme perturbations exceed this buffering capacity, cryptic genetic 
variation can be rapidly expressed, enabling phenotypic innovation. This 
mechanism -- where accumulated mutations remain phenotypically silent until 
environmental stress or genetic perturbation releases them -- may explain 
evolutionary transitions between fitness peaks without requiring intermediate 
forms of reduced fitness. Canalization is thus a key concept in evolutionary biology, forming a cornerstone of natural selection and the emergence of new phenotypes~\cite{von2011physics,hallgrimsson2019developmental}. It is also crucial for understanding developmental and physiological processes that shape disease outcomes in animals and influence crop yields in plants.
For instance, transitions to new phenotypes have been implicated as one of the driving forces behind tumorigenesis~\cite{ashworth2011genetic,jia2017phenotypic}.

To translate these qualitative ideas about canalization and robustness into a quantitative setting, it is necessary to adopt a mathematical framework that captures the logic and dynamics of gene regulation. Discrete dynamical systems, most prominently Boolean networks, provide the ideal framework for studying canalization because they explicitly represent the logical structure of regulatory interactions, the very substrate upon which canalization operates.

\section{Discrete dynamical gene regulatory network models}\label{sec:dds}
GRNs can be modeled as discrete dynamical systems,  e.g., Boolean networks (originally introduced by Kauffman~\cite{kauffman1969metabolic} and recently reviewed by Schwab et al.~\cite{schwab2020concepts}) or generalized multistate versions. Discrete models are intuitive, and simple to describe and analyze. They have therefore become an increasingly popular modeling framework for the study of GRNs. More than 150 discrete GRN models underlying a multitude of processes in single-celled organisms, plants, and animals have been generated, $>$80\% since 2012~\cite{kadelka2024meta}. Discrete models yield qualitative results even when quantitative kinetic parameters are unavailable, as is frequently the case in biology. While continuous differential equation models harbor the potential for quantitative predictions, their high parameter count makes them difficult to fit reliably to sparse data ~\cite{lahdesmaki2003learning,karlebach2008modelling}. However, an accurate formulation of a discrete model still requires substantial experimental data~\cite{lee2009computational,pratapa2020benchmarking}. For this reason, almost all discrete GRN models published thus far focus on specific biological modules containing only those genes and proteins directly involved in the process of interest~\cite{kadelka2023modularity}.

\begin{figure}
    \centering
    \includegraphics[width=\linewidth]{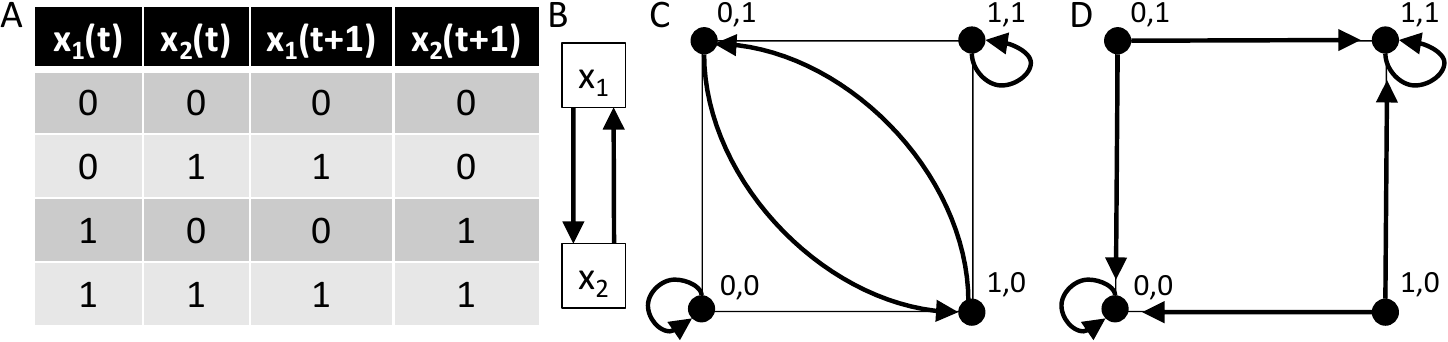}
    \caption{{\bf Example of a 2-node Boolean network $F(x_1,x_2) = (x_2,x_1)$.} (A) Wiring diagram indicating that $x_1$ and $x_2$ regulate each other. (B) Boolean update rules in truth table format, containing one row for each state of the network. (C) Deterministic synchronous state transition graph, containing three attractors: two steady states and a 2-cycle. (D) Stochastic asynchronous state transition graph, containing the same two steady states but no cyclic attractor.}
    \label{fig:example_bn}
\end{figure}

Mathematically, a \emph{discrete dynamical system} in the variables $x_1, \ldots, x_n$ is a function 
$$F = (f_1,\ldots, f_n) : \Ff^n \to \Ff^n,$$
where the Cartesian product $\Ff^n = \Ff \times \cdots \times \Ff$ defines the \emph{state space}, while each $f_i : \Ff^n \to \Ff$ specifies an \emph{update function} or \emph{update rule} that describes the future value of $x_i$ given the present value of all variables, i.e., $f_i$ captures the underlying regulatory logic. The set $\Ff$ contains all possible expression values of the variables. If $|\Ff| = 2$, e.g., $\Ff = \{0,1\}$, then $F$ is a \emph{Boolean network}, with $0$ and $1$ corresponding to an unexpressed and expressed gene or gene product, respectively. See Fig.~\ref{fig:example_bn}A for an example. Two directed graphs can be associated with $F$: 
\begin{enumerate}[(i)]
    \item The \emph{wiring diagram} (or \emph{dependency graph}) contains one node for each variable $x_i$, and has a directed edge from $x_i$ to $x_j$ if the update rule $f_j$ depends on $x_i$ (Fig.~\ref{fig:example_bn}B).
    \item The \emph{state transition graph} contains as nodes all possible states $\mathbf x = (x_1,\ldots,x_n)\in \Ff^n$. Under a synchronous updating scheme, all nodes are updated simultaneously and the state transition graph is deterministic with a directed edge from $\mathbf x$ to $\mathbf y$ if $F(\mathbf x) = \mathbf y$ (Fig.~\ref{fig:example_bn}C). Under asynchronous updating schemes, a single node is typically updated at a time, enabling the implementation of different time scales and yielding a non-deterministic state transition graph~(Fig.~\ref{fig:example_bn}D)~\cite{thomas1990biological}. 
\end{enumerate}

Due to the finite size of the state space, any state eventually transitions to an \emph{attractor}. Under synchronous update, the attractor is either a \emph{steady state} (also known as \emph{fixed point}) or a \emph{limit cycle}, while under asynchronous update, it is either a steady state or a \emph{trap space}~\cite{hopfensitz2013attractors}. 
Biologically, attractors represent self-maintaining regulatory states that cells settle into and stably occupy. In development, different attractors correspond to differentiated cell types, each characterized by a stable gene expression profile~\cite{kauffman1993origins}. In disease models, 
attractors can represent healthy versus pathological phenotypes. For instance, a Boolean GRN model of large granular lymphocyte leukemia exhibits two attractors: a ``healthy" phenotype corresponding to cell death of activated T cells and a ``diseased" phenotype corresponding to abnormal survival~\cite{zhang2008network,saadatpour2011dynamical}. An analysis of the attractor landscape (i.e., which states transition to which attractors) is thus key to understanding the dynamics of GRNs~\cite{Davidich08,choi2012attractor}, as well as for active control of the dynamics through drugs targeting specific gene interactions~\cite{campbell2019edgetic}.

All update rules can be represented as polynomials over a finite field, yielding a polynomial dynamical system and enabling the application of algebraic geometry techniques, e.g., for the efficient identification of steady states~\cite{laubenbacher2004computational,veliz2010polynomial}. Moreover, discrete dynamical systems can be viewed as Markov chains, allowing the use of tools from Markov theory~\cite{xiao2009tutorial}. This modeling framework, with its explicit representation of regulatory logic through update functions, provides the foundation for formalizing and quantifying canalization, as I describe next.

\section{The mathematical theory behind canalization}
\begin{figure}
    \centering
    \includegraphics[width=0.7\linewidth]{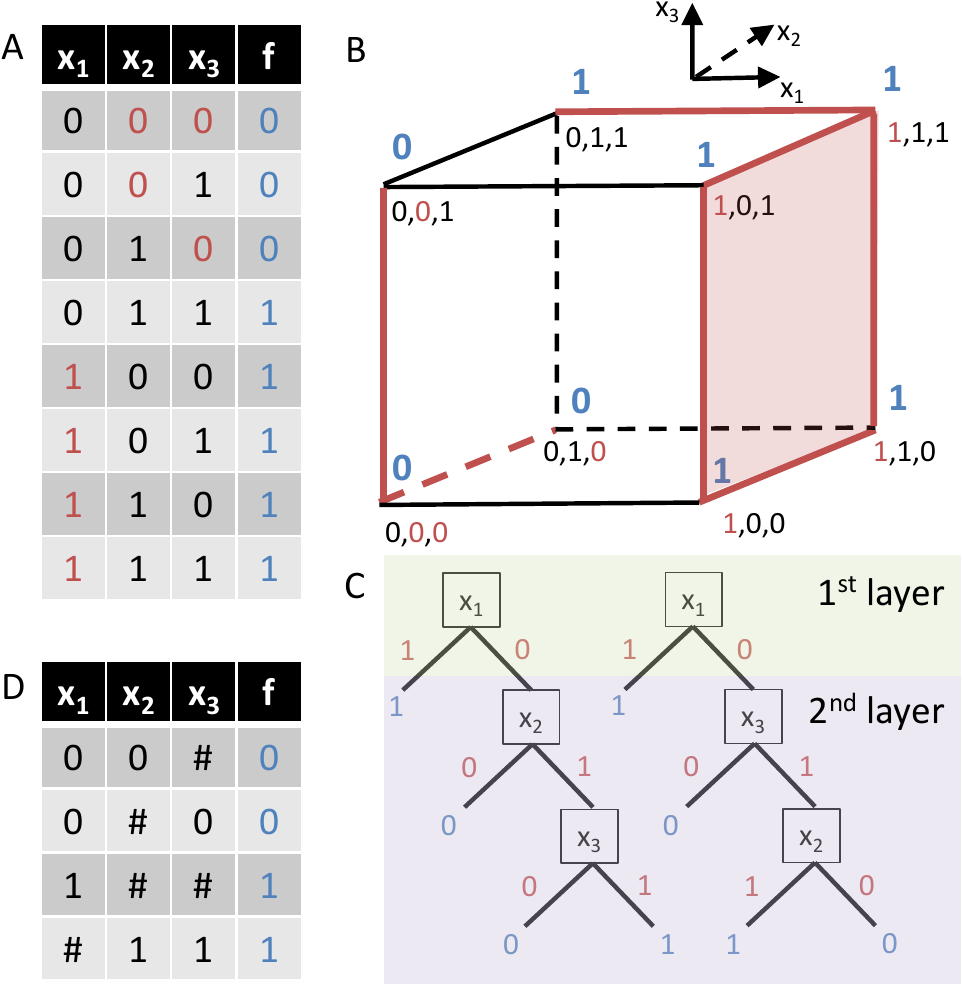}
    \caption{{\bf Example of a Boolean nested canalizing function.} 
    (A) Truth table of the 3-input Boolean NCF $f(x_1,x_2,x_3) = x_1 \vee (x_2 \wedge x_3)$. Setting $x_1=1$ canalizes $f$ to $1$. Setting $x_2=0$ or $x_3=0$ canalizes the subfunction $f(x_1=0,x_2,x_3) = x_2 \wedge x_3$. This means $f$ is an NCF with layer structure $(1,2)$.
    (B) Canalizing properties can be derived from a Boolean (hyper)cube labeled according to $f$. The proportion of ($n-k$)-dimensional faces that are constant is the $k$-set canalizing proportion. For example, the constant (red) face $(1,\#,\#)$ indicates that $x_1=1$ canalizes $f$. Similarly, the constant (red) edges $(0,0,\#)$ and $(0,\#,0)$ indicate that $x_2=0$ or $x_3=0$ independently canalize the subfunction. 
    (C) Two possible nested evaluation trees, highlighting that $x_1$ is in the most important canalizing layer and that $x_2$ and $x_3$ are equally important.
    (D) Reduced truth table where \# indicates that a certain input does not matter. The edge effectiveness and input redundancy are computed from this reduced table.}
    \label{fig:example}
\end{figure}

Shortly after introducing Boolean networks as models of GRNs~\cite{kauffman1969metabolic}, Stuart Kauffman proposed Boolean canalizing functions as a suitable class of update rules~\cite{kauffman1974large}. A  function $f: \{0,1\}^n \rightarrow \{0,1\}$ is \emph{canalizing} if it has at least one input variable $x_i$ (called a canalizing variable) such that setting $x_i$ to a specific value $a \in \{0,1\}$ (the canalizing input) fully determines the output to be $b \in \{0,1\}$ (the canalized output), regardless of all other inputs. The requirement that $f$ takes other values when $x_i\neq a$ ensures that constant functions are not considered canalizing~\cite{he2016stratification}. If the first variable does not take on the canalizing input value but there is a second variable with this canalizing property, the function is 2-canalizing. If k variables follow this pattern, the function is k-canalizing~\cite{he2016stratification}, and the number of variables that follow this pattern is known as \emph{canalizing depth}~\cite{layne2012nested}. If all variables follow this pattern (i.e., if the canalizing depth equals the number of inputs $n$), $f$ is a \emph{nested canalizing function} (\emph{NCF})~\cite{kauffman2003random}. Fig.~\ref{fig:example} shows an example of an NCF: $f(x_1,x_2,x_3) = x_1 \vee (x_2 \wedge x_3)$. 

\begin{figure}
    \centering
    \includegraphics[width=\linewidth]{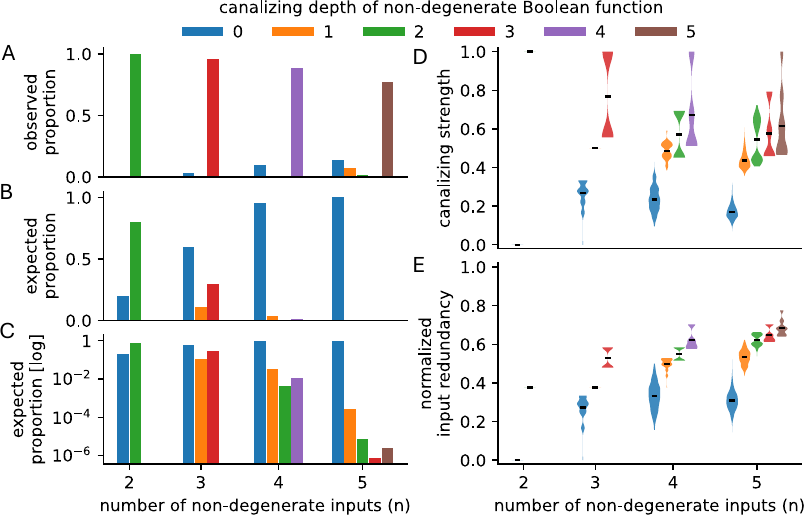}
    \caption{{\bf Relationships between canalization metrics for all non-degenerate 
Boolean functions with 2-5 inputs}. (A) Observed proportion of functions 
with specific canalizing depth in 122 expert-curated Boolean GRN models~\cite{kadelka2024meta}. (B,C) Proportion of non-degenerate functions 
with each canalizing depth, shown on (B) linear and (C) log scales, computed using exact formulas~\cite{kadelka2025number}. As the 
number of inputs increases, functions with high canalizing depth become 
exponentially rarer. (D,E) Distribution of (D) canalizing strength and (E) normalized input redundancy for non-degenerate functions with specific canalizing depth, generated using \texttt{BoolForge}~\cite{kadelka2025boolforge} and 1000 random functions per distribution. The high correlations between canalizing depth, canalizing strength, and input redundancy indicate that these metrics capture related but not identical aspects of canalization.}
    \label{fig:number_canalizing}
\end{figure}

Expert-curated Boolean GRN models are almost exclusively composed of canalizing or even nested canalizing functions~\cite{harris2002model,daniels2018criticality,kadelka2024meta}, underscoring the central role of canalization in gene regulation~(Fig.~\ref{fig:number_canalizing}A). The probability that an $n$-input Boolean or multistate function is canalizing or nested canalizing can be derived analytically~\cite{just2004number,li2013boolean,he2016stratification,kadelka2017multistate,dimitrova2022revealing,kadelka2025number}. As the number of variables increases, canalization -- and in particular the presence of multiple canalizing variables -- becomes an increasingly rare property~(Fig.~\ref{fig:number_canalizing}B,C), making the empirical prevalence of canalizing logic in biological systems all the more remarkable.

Every Boolean function can be expressed in a unique \emph{standard monomial form}, in which variables are partitioned into distinct \emph{layers} according to their dominance~\cite{he2016stratification}. For example, the NCF $f(x_1,x_2,x_3) = x_1 \vee (x_2 \wedge x_3)$ possesses two layers: the first layer contains the canalizing variable $x_1$ (since $f(x_1=1,x_2,x_3) = 1$), while the second layer contains $x_2$ and $x_3$ (since $f(x_1=0,x_2=1,x_3) = f(x_1=0,x_2,x_3=1) = 1$). Variables in the first layer are directly canalizing; those that become canalizing once all variables in the first layer have received their non-canalizing input belong to the second layer, and so forth. Variables that never become canalizing constitute the \emph{core polynomial}. The \emph{layer structure} specifies the number of variables in each layer, e.g., $(1,2)$ for the function $f$~\cite{kadelka2017influence}. NCFs with identical layer structure exhibit the same dynamical properties, such as the same \emph{average sensitivity}, defined as the probability that the function’s output changes when a single, randomly chosen input is flipped~\cite{kadelka2017influence}. The standard monomial form, and in particular the layer structure, therefore provides a principled framework for further classifying Boolean functions. Notably, published Boolean GRN models are enriched for NCFs with low average sensitivity~\cite{kadelka2024meta}, suggesting that evolution may favor such regulatory logic to enhance network stability.

More recently, the notion of \emph{collective canalization} has been introduced, extending the traditional variable-centric definition to a function-centric one~\cite{Reichhardt}. Rather than asking whether individual variables can determine the output, this framework asks whether subsets of variables collectively canalize the function. Formally, a Boolean $n$-input function is \emph{$k$-set canalizing} if there exists a subset of $k$ variables whose values, once specified, fully determine the output, regardless of the remaining $n-k$ inputs~\cite{kadelka2023collectively}. Under this definition, 1-set canalizing functions correspond exactly to classical canalizing functions. Further, for any $n\geq 2$, only two Boolean functions are not $(n-1)$-set canalizing: the parity function (XOR) and its complement (XNOR). The \emph{$k$-set canalizing proportion} quantifies what fraction of all $k$-variable subsets can collectively canalize a function, and the \emph{canalizing strength} -- a weighted average of these proportions -- describes the overall degree of canalization~(Fig.~\ref{fig:example}B)~\cite{kadelka2023collectively}. This strength equals one for maximally canalizing functions (single-layer NCFs where all variables are canalizing, such as AND or OR) and zero for minimally canalizing functions (XOR and XNOR, where knowledge of all variables is always required).

An alternative, information-theoretic perspective connects canalization to redundancy~\cite{marques2013canalization}. From this viewpoint, a highly canalized function contains redundant information: knowing one variable's value often makes other variables irrelevant~(Fig.~\ref{fig:example}D). This intuition can be formalized using the Quine-McCluskey minimization algorithm~\cite{mccluskey1956minimization}, which yields the redundancy associated with each input and its complement, termed \emph{edge effectiveness}~\cite{gates2021effective}. Summing all edge effectiveness values yields a Boolean function's \emph{effective degree}, whereas the total redundancy defines its \emph{input redundancy}. The resulting \emph{effective graph} is an edge-weighted version of the wiring diagram, where each edge weight reflects the effectiveness of that regulatory connection.

These diverse mathematical formalizations provide complementary lenses for quantifying canalization in Boolean functions. While highly correlated (Fig.~\ref{fig:number_canalizing}D,E), they emphasize different aspects: canalizing depth and layer structure focus on variable dominance hierarchies, canalizing strength on collective buffering, and input redundancy on information compression. The proliferation of these metrics reflects both the richness of the canalization concept and ongoing efforts to identify which formalization best captures the biological essence of robustness in gene regulation. I elaborate on this question next.

\section{Many definitions of canalization: which one is ``right"?}
The various definitions of canalization capture related, but not identical, aspects of regulatory robustness. As Fig.~\ref{fig:number_canalizing}D,E illustrates, functions with greater canalizing depth tend to exhibit higher canalizing strength and input redundancy. This correlation is expected: mechanisms that render certain variables dominant (high depth) often introduce functional redundancy (high input redundancy) and allow subsets of variables to determine the output (high collective canalization). Yet the correlations are not perfect, indicating that these metrics emphasize different structural features of regulatory logic.

\begin{figure}
    \centering
    \includegraphics[width=0.4\linewidth]{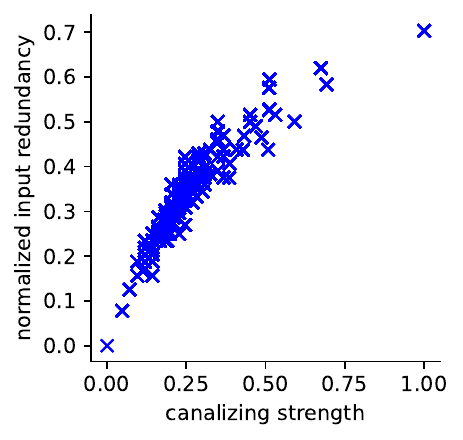}
    \caption{{\bf Difference between canalizing strength and normalized input redundancy.} All 4-input non-degenerate Boolean functions were generated and analyzed using \texttt{BoolForge}~\cite{kadelka2025boolforge}.}
    \label{fig:collective_canalization}
\end{figure}

For example, consider two 5-input Boolean functions. The function $f$ outputs $1$ whenever $x_1=1$ or when an odd number of the remaining four inputs are $1$. The function $g$ outputs $1$ whenever at least two of its five inputs are $1$. The variable $x_1$ canalizes $f$, whereas no single variable canalizes $g$. Nevertheless, $g$ has higher canalizing strength ($0.426$) and normalized input redundancy ($0.525$) than $f$ ($0.336$ and $0.375$, respectively). This example shows that canalizing depth captures variable-level dominance but may overlook distributed forms of robustness quantified by the other two metrics. Similarly, a comprehensive enumeration of all non-degenerate 4-input Boolean functions reveals the same pattern: despite high overall correlation ($\rho_{\text{Spearman}} = 0.958$), some functions have high canalizing strength but relatively low input redundancy and vice versa (Fig.~\ref{fig:collective_canalization}).

These observations motivate a fundamental question: which definition of canalization is biologically most meaningful? The answer likely depends on the biological context and the specific question being asked.
\begin{itemize}
    \item For understanding evolutionary constraints, canalizing depth and layer structure may be most relevant, as they directly relate to how easily mutations in different genes can alter phenotypes. Genes in dominant layers are more ``evolutionarily constrained" because mutations affecting them have larger phenotypic consequences~\cite{kadelka2017influence}.
    \item For control and therapeutic intervention, input redundancy and the effective graph framework may be most useful, as they identify which regulatory connections are functionally important versus redundant~\cite{gates2021effective}. These insights can guide strategies for driving transitions between attractors, such as in disease intervention~\cite{campbell2019edgetic,parmer2022influence}.
    \item For comparing biological networks to random ensembles, canalizing strength and normalized input redundancy both provide a single, normalized metric (ranging from 0 to 1)~\cite{kadelka2023collectively,gates2021effective}. This facilitates statistical comparisons across networks with different sizes and structures.
\end{itemize}
Rather than seeking a single ``correct" definition, it is more productive to view canalization as a multifaceted property. Just as graph theory includes multiple centrality measures -- such as degree, betweenness, eigenvector centrality -- each capturing a different notion of node importance -- canalization encompasses several correlated but distinct dimensions of regulatory robustness.

A deeper question is whether any of these mathematical notions fully align with Waddington's original concept of canalization. Waddington emphasized a developmental system’s ability to maintain a normal trajectory despite perturbations~\cite{waddington1942canalization}, highlighting dynamical robustness at the network level rather than structural properties of individual functions. Recent work linking function-level canalization to network-level stability (Section~\ref{sec:stability}) helps bridge this gap, but the ultimate test is empirical: Which canalization metrics best predict phenotypic buffering in real biological systems? Addressing this question will require systematic, large-scale comparisons that integrate computational predictions with experimental measurements of robustness.

\section{Canalization and its role in conferring stability to GRNs}\label{sec:stability}
The extensive body of work inspired by Kauffman’s pioneering work has firmly established that Boolean networks governed by canalizing update rules exhibit enhanced dynamical stability: they possess fewer attractors, shorter limit cycles, and reduced sensitivity to perturbations compared to networks with random update rules~\cite{kauffman2004genetic,Shmul04,Karlsson07,kadelka2017influence}. In this section, I synthesize key results linking function-level canalization to network-level robustness, with emphasis on the different stability metrics and their biological interpretation.

\subsection*{Dynamical regimes and the criticality hypothesis}
A useful starting point for understanding how canalization shapes dynamics is the classification of Boolean networks into \emph{ordered, critical, and chaotic regimes}. Derrida and Pomeau’s seminal framework~\cite{Derrida1,Derrida2} characterizes these regimes by how a small perturbation propagates through the system. The \emph{Derrida value} (or Derrida coefficient) measures the expected number of node-state changes after one update step, given an initial single-node perturbation: values less than 1 correspond to the ordered regime (perturbations decay), values greater than 1 to the chaotic regime (perturbations amplify), and values near 1 to the critical regime.

The \emph{criticality hypothesis} proposes that biological GRNs operate near this boundary between order and chaos, where systems can remain robust yet still respond adaptively to external signals~\cite{balleza2008critical,roli2018dynamical}. Indeed, many published Boolean GRN models exhibit mean average sensitivities close to 1~\cite{daniels2018criticality,kadelka2024meta}. However, recent work has revealed that these models include a substantial number of source nodes -- variables representing external inputs or cellular context with no incoming edges -- which artificially inflate average sensitivity. Accounting properly for these source nodes shifts most models toward the ordered regime~\cite{park2023models}, suggesting that biological networks may be more stable than previously thought.

A classical mean-field approximation provides a simple criterion for the transition between order and chaos~\cite{Derrida2}: a Boolean network is expected to be critical when  
$
2\langle k\rangle\, p(1-p) = 1,
$
where $\langle k\rangle$ is the average in-degree and $p$ is the average bias of the update rules. This expression captures the intuition that networks with high connectivity or balanced outputs (i.e., $p\approx 0.5$) are more prone to chaotic dynamics. However, this threshold relies on nominal connectivity and does not account for the reduction in \emph{effective} regulatory influence introduced by canalization. Recent work has refined this criterion by replacing $\langle k\rangle$ with the \emph{effective degree} $\langle k_e\rangle$, derived from the information-theoretic definition of canalization as the sum of edge effectiveness values~\cite{manicka2022effective}. The resulting improved condition,
$
3.94\,\langle k_e\rangle\, p(1-p) = 1,
$
more sharply distinguishes ordered from chaotic networks and reveals that many biological GRNs lie deeper in the ordered regime than predicted by the classical approximation.

\subsection*{How canalization modulates stability}
Canalization modulates stability primarily through its effect on update function sensitivity. NCFs with the same layer structure share the same average sensitivity~\cite{kadelka2017influence}, and this sensitivity varies systematically across layer structures. Highly biased NCFs (e.g., single-layer AND or OR functions) exhibit low sensitivity, whereas less biased NCFs show higher sensitivity. Remarkably, Boolean networks composed entirely of NCFs (with random layer structure) have expected Derrida values equal to 1 -- independent of each function’s in-degree -- placing such networks exactly at criticality. In contrast, networks with random canalizing or entirely random update functions become increasingly chaotic as connectivity increases~\cite{shmulevich2004activities}.

Published Boolean GRN models are strongly enriched for NCFs, particularly those with low sensitivity~\cite{kadelka2024meta}. One interpretation is that evolution preferentially selects insensitive NCFs to counterbalance more sensitive update rules elsewhere in the network, thereby maintaining dynamics near criticality. This “compositional tuning’’ suggests that criticality may emerge from how stabilizing and destabilizing update rules are mixed across the network rather than from a uniform choice of function type.

\subsection*{Coherence and the stability of network attractors}
The Derrida value can be computed from the average sensitivities of all update rules~\cite{kadelka2017influence}. It does not depend on the underlying wiring diagram, and thus fails to capture the influence of feedback loops and other topological features that may affect network stability~\cite{squires2014stability}. To address this limitation, multi-timestep Derrida values have been considered, measuring long-term divergence between trajectories that begin with a one-bit difference. However, such measures are sensitive to phase shifts: even if two trajectories converge to the same limit cycle, they may never coincide in state space if they enter the cycle at different time points~\cite{park2023models}.

Two complementary metrics avoid this issue. The \emph{fragility} quantifies the long-term average difference between two initially perturbed trajectories~\cite{park2023models}. The \emph{coherence} measures the probability that two states differing by a single bit transition to the same attractor, defined for synchronous~\cite{willadsenwiles} and stochastic updates~\cite{kadelka2013stabilizing}, and as quasicoherence for asynchronous updates~\cite{park2023models}. Networks with high coherence have robust basins of attraction, meaning small perturbations rarely induce phenotype switches.

\begin{figure}
    \centering
    \includegraphics[width=\linewidth]{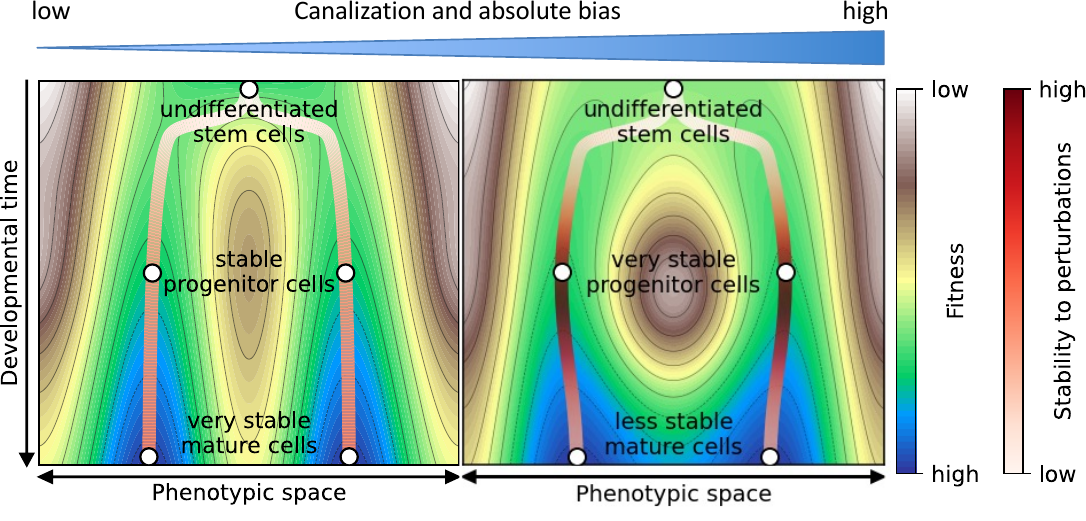}
    \caption{{\bf Canalization creates a stability gradient within Waddington's landscape.}
High canalization deepens the valleys that channel developing cells toward robust phenotypic outcomes (left–right gradient, top bar), reinforcing the stability of developmental trajectories. However, when multiple attractors in highly canalized networks coexist, they are concentrated in nearby regions of the state space. This crowding flattens the local landscape near attractors relative to mid-trajectory regions, leaving terminal cell states positioned closer to basin boundaries. As a consequence, mature phenotypes can be more susceptible to directed or coordinated perturbations even when the overall developmental funnel is highly robust. This ``intra-valley stability gradient" offers a mechanistic interpretation of how canalization can simultaneously ensure reproducible development and permit regulated phenotype switching in contexts such as reprogramming, regeneration, and pathological transitions.
    }
    \label{fig:waddington}
\end{figure}

Canalization strongly increases coherence: networks governed by canalizing update rules, such as most expert-curated Boolean GRN models, display significantly higher coherence than random networks on the same wiring diagram~\cite{bavisetty2025attractors}. A recent refinement distinguishes \emph{basin coherence} (robustness of all states in a basin) from \emph{attractor coherence} (robustness of attractor states themselves)~\cite{bavisetty2025attractors}. Kauffman already described the measure of attractor coherence in his pioneering work but it was never formally analyzed~\cite{kauffman1969metabolic}. Paradoxically, canalization reduces the relative stability of attractors compared to their basins~\cite{bavisetty2025attractors}. This arises from functional bias: canalizing functions tend to be biased toward 0 or 1 outputs~\cite{shmulevich2003role}, which confines attractors to similar regions of the state space~\cite{socolar2003scaling}, making transitions between them more likely. The result is a subtle trade-off: canalization stabilizes developmental trajectories (high overall basin coherence) but may render terminal phenotypes more susceptible to targeted perturbations (lower attractor coherence). This refines Waddington's landscape metaphor (Fig.~\ref{fig:waddington}) and has important implications for understanding processes like stem cell reprogramming, wound healing, and disease-related transitions such as cancer differentiation~\cite{bavisetty2025attractors}.

\section{Disentangling correlated properties}\label{sec:disentangling}
A fundamental challenge in understanding GRN design principles is that many structural and dynamical features are strongly correlated, complicating causal inference. Biological networks exhibit high canalization, low connectivity, functional bias, and near-critical dynamics~\cite{shen2002network,macneil2011gene,deritei2016principles,gorochowski2018organization,daniels2018criticality, kadelka2024meta, kadelka2024canalization}. Because these properties co-occur, it remains unclear which features evolution actively selects for versus which arise as byproducts of selection on other traits.

For example, consider the relationship between canalization and bias. Canalizing functions, particularly the most canalizing NCFs with single-layer structure (e.g., AND, OR), are necessarily highly biased, i.e., they output 0 or 1 with high probability~\cite{shmulevich2003role}. Is the observed functional bias in biological networks a direct target of selection, or does it emerge automatically because evolution favors canalization? Similarly, modularity (the decomposition of networks into semi-independent subnetworks) may arise from selection for evolvability, or it may be a byproduct of selection for specific dynamical behaviors that happen to require modular architectures~\cite{espinosa2018role,kadelka2023modularity}.

\subsection*{Toward causal understanding through comparative analysis}

Large-scale comparative analyses offer the most promising path toward disentangling these correlations. By analyzing hundreds of independently curated biological network models, researchers can identify which properties are universally present (possibly suggesting strong selection) versus which vary across biological contexts (possibly suggesting contingent evolution or byproduct status). However, such meta-analyses come with inherent flaws. Despite progress, the number of published, experimentally validated GRN models remains relatively small~\cite{helikar2012cell,kadelka2024meta}, and almost all models are for model organisms, potentially biasing meta-analyses. Moreover, researchers may be more likely to publish models with ``interesting" properties (e.g., near criticality, high canalization), or, as in~\cite{hinkelmann2012inferring,zhou2016relative}, they may even use putative design principles such as nested canalization to constrain the difficult network inference problem. This may substantially inflate the apparent prevalence of these features. Addressing these challenges will require sustained collaboration between theoreticians, experimentalists, and computational biologists. The payoff, a mechanistic understanding of how molecular regulatory logic produces robust phenotypes, would represent a major advance in systems biology.

A complementary approach is computational experimentation: systematically varying one network property while holding others constant, then measuring effects on metrics of interest. For example, the recently postulated high ``approximability" of biological networks~\cite{manicka2023nonlinearity}, i.e., the ability to accurately approximate biological network dynamics by simpler models, is almost entirely explained by their abundance of canalization~\cite{kadelka2024canalization}. This finding suggests that approximability is not a direct target of selection but rather emerges from canalization. Such studies provide a template for future work disentangling other correlated properties.
\texttt{BoolForge}, a recently developed Python package for the random generation and analysis of Boolean functions and networks (with a particular focus on canalization), greatly facilitates the design of such computational studies~\cite{kadelka2025boolforge}.

\subsection*{The evolutionary origins of canalization}

Understanding why canalization is so prevalent in gene regulation requires evolutionary modeling. 
Several hypotheses have been proposed:

\begin{enumerate}
    \item Direct selection for robustness: Canalization may be favored because robust phenotypes have higher fitness, particularly in fluctuating environments~\cite{von2011physics,hallgrimsson2019developmental}. Organisms that maintain proper development despite environmental perturbations or minor mutations out-compete less robust individuals.
   \item Selection for evolvability: Canalization enables the accumulation of cryptic genetic variation that can be released under stress, facilitating rapid adaptation to new environments~\cite{ashworth2011genetic,jia2017phenotypic}. Networks that buffer most mutations while permitting occasional large phenotypic shifts may be evolutionarily advantageous.
   \item Developmental constraint: The biochemical mechanisms underlying gene regulation (e.g., cooperative binding, allosteric regulation of transcription factors, and ultrasensitive threshold responses) naturally produce canalizing logic~\cite{kauffman1974large,ferrell2014ultrasensitivity}. Thus, evolution may not ``select for" canalization directly so much as operate within the constraints of available molecular mechanisms that inherently favor canalized behavior.
\end{enumerate}
These hypotheses are not mutually exclusive, and all may contribute to varying degrees to the prevalence of canalization. Distinguishing among them requires carefully designed computational evolutionary studies (see e.g.~\cite{siegal2002waddington,runneburger2016and,huizinga2018emergence}), as well as integrating mathematical modeling with experimental studies of evolution, an important frontier for future research.

\section{Beyond Boolean logic}

While many regulatory mechanisms can be accurately described in Boolean logic, this is not always the case. First introduced by Rene Thomas in 1991~\cite{thomas1991regulatory}, multistate network
models become increasingly common. Here, some (or all) variables take values in finite sets $\Ff$ with $|\Ff| > 2$ (e.g., 
$\Ff = 
\{0,1,2\}$ for ternary logic). For example, a regulatory network for the G1/S checkpoint pathway includes the ternary input node DNA damage (0: no damage, 1: low damage, 2: high irreparable damage) as well as ternary internal nodes such as the tumor suppressor p53, which is inhibited in the absence of DNA damage, induced in its presence, and at high concentration leads to cell apoptosis~\cite{issler2017microrna}. Similarly, many genes in a model of the innate immune response to ischemic injury are ternary indicating either low, intermediate, and high expression levels or inactive, active and hyperactive proteins~\cite{dimitrova2018innate}. Moreover, proteins can be phosphorylated at multiple sites, creating more than two functional states~\cite{salazar2009multisite}. Some inputs (i.e., cellular contexts) and outputs (i.e., phenotypes) are also naturally multi-valued. This motivates the need for multistate discrete models and raises important questions: How can canalization be defined for multistate functions? Do the stability results derived for Boolean networks extend to multistate systems? 

\subsection*{Defining multistate canalization}
\begin{figure}
    \centering
    \includegraphics[width=0.7\linewidth]{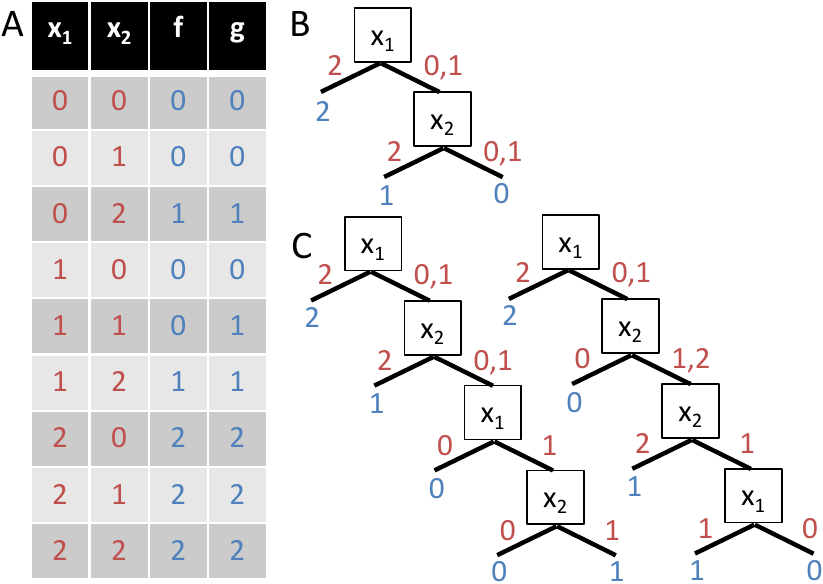}
    \caption{{\bf Different definitions of nested canalization in the multistate setting.} (A) Example of a nested canalizing ternary function $f$ and a weakly nested canalizing ternary function $g$ in truth table format (inputs: red, outputs: blue), (B) Unique shortest decision diagram of $f$, (C) Two of many shortest decision diagrams of $g$.}
    \label{fig:multistate}
\end{figure}

Kauffman’s traditional definition of canalization extends naturally to the multistate setting. Instead of a single canalizing input value, a multistate canalizing function $f: \Ff^n\to \Ff$ may possess a \emph{canalizing input segment} $S\subset \Ff$ with $1\leq |S| < |\Ff|$. Typically, $S$ is also required to contain exactly one of the endpoints of $\Ff$~\cite{murrugarra2011regulatory}. The notion of nested canalization can also be generalized in an iterative way (Fig.~\ref{fig:multistate}A,B). As an example, consider the ternary function
$$f(x_1,x_2)=\begin{cases}
    2 & \ \text{if}\ x_1 = 2,\\
    1 & \ \text{if}\ x_1 \neq 2\ \text{and}\ x_2 = 2,\\
    0 & \ \text{otherwise}.
\end{cases}$$
Closed-form expressions for the abundance and the expected average sensitivity of such multistate NCFs, as well as Derrida values for networks governed by them, have been derived~\cite{kadelka2017multistate}. 

However, there are alternative ways to define multistate nested canalization. This is due to the fact that a ``canalizing variable" may need to be considered multiple times when evaluating a multistate function. Consider the ternary function 
$$g(x_1,x_2)=\begin{cases}
    2 & \ \text{if}\ x_1 = 2,\\
    1 & \ \text{if}\ x_1 \neq 2\ \text{and}\ x_1 + x_2 \geq 2,\\
    0 & \ \text{otherwise}.
\end{cases}$$
Here, setting $x_1=2$ canalizes $g$ (Fig.~\ref{fig:multistate}A,C), yet the subfunction that needs to be evaluated when $x_1\neq 2$ still depends on $x_1$. Because $g$ can nevertheless be evaluated in a nested fashion, it is natural to consider it nested canalizing. To distinguish this behavior from the stricter case where each variable appears only once, such functions are termed weakly nested canalizing~\cite{remy2023average}. Every nested canalizing function is trivially weakly nested canalizing, but not vice versa.

The existence of weakly nested canalizing functions complicates the layer structure concept. Unlike the Boolean case, where the standard monomial form uniquely assigns each variable to a single layer, multistate functions do not admit a comparably clean decomposition~\cite{kadelka2017multistate,murrugarra2021quantifying}. A variable may canalize the output for some input values yet interact non-canalizingly with other variables for others, preventing a unique layered representation. Consequently, canonical metrics such as canalizing depth and layer structure do not generalize directly to multistate functions.

An alternative perspective based on collective canalization appears to generalize more naturally. The definition of k-set canalization -- whether a subset of k variables suffices to determine the output regardless of the remaining variables -- extends immediately to any finite-valued function~\cite{kadelka2023collectively}. The k-set canalizing proportion and canalizing strength can be computed in nearly the same way for multistate and Boolean functions, providing a unified framework across different state-space sizes. Future work should focus on sorting out mathematical details and providing formulas for statistical properties, such as the distribution of canalizing strength. Likewise, notions of input redundancy and the effective graph approach should also extend straightforwardly~\cite{marques2013canalization,gates2021effective}, as multi-valued versions of the Quine–McCluskey algorithm enable computation of edge effectiveness in multistate networks. As in the Boolean case, it remains an open question whether one definition of multistate canalization is biologically most useful, or whether different definitions capture distinct aspects of the biological phenomenon of canalization.

\subsection*{Dynamics and stability of multistate networks}
The role of canalization in stabilizing Boolean networks has been studied extensively using a variety of stability measures. Most of these measures -- such as the Derrida value, fragility, or coherence -- quantify how a small perturbation propagates through the network and affects short- or long-term dynamics. In the Boolean setting, the difference between two network states is naturally defined via the Hamming distance, which counts the number of differing bits. Although this idea generalizes formally to multistate networks, it is far less clear what type of perturbation is biologically meaningful. For example, a ternary gene at a high expression level is more likely to fluctuate to an intermediate level than to jump directly to a low level. Should stability therefore be assessed only under “one-level’’ perturbations? If so, does a variable in an intermediate state experience twice the perturbation probability of one at an extreme? Modeling choices of this kind -- regarding both how perturbations are introduced and how distances among multistate values are defined -- directly shape the resulting stability assessments. In this sense, even the notion of a ``small perturbation" becomes model-dependent in the multistate setting, underscoring the need for perturbation metrics that more faithfully reflect biochemical variability.

\subsection*{Empirical multistate GRN models}
Despite the biological motivations outlined above, multistate GRN models remain rare compared to their Boolean counterparts. In assembling a repository of more than 150 expert-curated Boolean GRN models by parsing the entire PubMed abstract corpus~\cite{kadelka2024meta}, we identified only 18 multistate models. Because our keyword-based, semi-automated search was optimized for Boolean models, some multistate models were likely missed; nonetheless, the disparity is striking.

This scarcity reflects both practical and conceptual challenges. First, data requirements for multistate modeling are substantially higher. Inferring update rules with multiple expression levels requires fine-grained quantitative measurements rather than the simpler expressed/unexpressed distinctions sufficient for Boolean models. Second, experimental validation is more demanding: perturbations must achieve specific intermediate expression states rather than complete knockouts or overexpressions, which can be technically challenging. Notably, certain systems -- such as Arabidopsis root development, where controlled knockouts (0), knockdowns (1), and overexpression (2) of multiple genes can be achieved efficiently -- offer promising platforms for generating and validating multistate GRNs. Finally, Boolean logic maps naturally onto biologists' intuitions about gene ``activity," whereas multistate functions are required only when more nuanced regulatory mechanisms must be captured.

Analyses of expert-curated Boolean GRN models have been tremendously helpful in uncovering putative biological design principles~\cite{daniels2018criticality,kadelka2024meta}. The field would thus greatly benefit from a dedicated, comprehensive repository of all published multistate GRN models, analogous to existing Boolean model databases~\cite{helikar2012cell,kadelka2024meta,pastva2023repository}. Such a resource would enable systematic comparative analyses, and guide the development of improved inference and validation methodologies.

\section{Discussion}

Canalization provides a unifying perspective on how gene regulatory networks achieve robust function despite pervasive molecular noise, environmental variation, and genetic perturbation. Across decades of work -- from Waddington's developmental insights to modern discrete dynamical systems theory -- canalization has emerged as a central organizing principle of GRNs. The mathematical results reviewed here demonstrate that canalizing update rules sharply constrain network dynamics, promote ordered or near-critical behavior, and generate attractor landscapes with robust basins. At the same time, recent findings such as the coherence gap reveal that canalization produces subtle trade-offs: while developmental trajectories are stabilized, mature phenotypes may become more susceptible to targeted perturbations. Understanding when evolution exploits versus avoids such ``edge-of-stability" behavior remains an important open question. 

A recurring theme in this perspective is that canalization is multifaceted. Variable-centric, collective, and redundancy-based notions capture overlapping but distinct features, much like centrality measures in graph theory. Rather than seeking a single correct metric, future work should aim to understand which forms of canalization map most closely onto experimentally observed robustness. This requires systematic empirical studies linking quantitative canalization metrics to phenotypic buffering \textit{in vivo} -- studies that are still very rare. Large-scale comparative analyses, meta-analyses of curated GRN models, and perturbation-response experiments in systems where intermediate expression levels can be controlled will be crucial in bridging theory and experiment.

The extension to multistate regulatory logic remains one of the least explored but most biologically significant frontiers. Many biochemical interactions -- such as multisite phosphorylation and multi-level cellular contexts -- are inherently multivalued. Yet multistate canalization theory is still in its infancy. The existence of weakly nested canalizing functions, the lack of a canonical layer structure, and the dependence of stability on how perturbations are defined all point to conceptual challenges distinct from the Boolean case. Addressing them will require new mathematical tools, improved perturbation metrics that reflect biochemical noise, and curated repositories of multistate GRNs. These resources would enable clearer comparisons between Boolean and multistate dynamics and help determine when more states are necessary for biological realism.

Looking forward, canalization offers a powerful conceptual bridge between the logic of gene regulation and the dynamics of phenotype. Its relevance spans molecular biology, evolution, control theory, and network inference. Advancing the field will require integrating algebraic methods, computational experiments, evolutionary modeling, and targeted laboratory validation. By developing a unified theory of canalization in both Boolean and multistate systems -- and by grounding that theory in empirical data -- we can move closer to a mechanistic understanding of how regulatory architectures produce the remarkable stability and adaptability observed across living systems.

\subsection*{Data availability}
Not applicable. 

\subsection*{Code availability}
All code used to perform the analyses and create the figures is available at \href{https://github.com/ckadelka/PerspectiveCanalization}{https://github.com/ckadelka/PerspectiveCanalization}.


\subsection*{Acknowledgements}
The author acknowledges support from the Simons Foundation (grant 712537) and the National Science Foundation (grants DMS-2424632 and DMS-2451973).


\subsection*{Competing interests.}
The author declares no competing interests.

\end{document}